# Vega library for processing DICOM data required in Monte Carlo verification of radiotherapy treatment plans

Christopher Locke[1] and Sergei Zavgorodni[1,2]

[1] Department of Physics and Astronomy, University of Victoria, Victoria BC, Canada.
[2] Department of Medical Physics, British Columbia Cancer Agency, Vancouver Island Center, Victoria BC, Canada.

**Running title: DICOM library for MC calculations**

Corresponding author:    Sergei Zavgorodni,
Department of Medical Physics
BC Cancer Agency, Vancouver Island Centre
2410 Lee Street, Victoria, BC, Canada, V8R 6V5
Phone: +1 250-519-5628
Fax: +1 250-519-2024
szavgorodni@bccancer.bc.ca

## Abstract

Monte Carlo (MC) methods provide the most accurate to-date dose calculations in heterogeneous media and complex geometries, and this spawns increasing interest in incorporating MC calculations into treatment planning quality assurance process. This involves MC dose calculations for clinically produced treatment plans. To perform these calculations, a number of treatment plan parameters specifying radiation beam and patient geometries need to be transferred to MC codes, such as BEAMnrc and DOSXYZnrc. Extracting these parameters from DICOM files is not a trivial task, one that has previously been performed mostly using Matlab-based software. This paper describes the DICOM tags that contain information required for MC modeling of conformal and IMRT plans, and reports the development of an in-house DICOM interface, through a library (named Vega) of platform-independent, object-oriented C++ codes. The Vega library is small and succinct, offering just the fundamental functions for reading/modifying/writing DICOM files in a C++ program. The library, however, is flexible enough to extract all MC required data from DICOM files, and write MC produced dose distributions into DICOM files that can then be processed in a treatment planning system environment. The library can be made available upon request to the authors.

## 1. Introduction

Calculation accuracy of the modern commercial radiotherapy treatment planning algorithms does not always meet modern dosimetric requirements in situations where dose distributions are delivered to volumes with heterogeneous media, such as air-tissue and lung-tissue interfaces[1-3]. It is well recognized that the most accurate methods to calculate the dose near such interfaces are based on Monte Carlo (MC) electron/photon particle transport algorithms. In this situation a number of group[2,4-10] implemented a process that enables re-calculation of treatment plans using MC based codes.

Many other groups are also in the process of, or interested in, implementing such MC based "calculation engines" in their institutions, and most "building blocks" required for this implementation are available or published. MC codes commonly used in these implementations are BEAM [11,12] and DOSXYZnrc[13,14]. They are utilized for modeling radiation transport through linac and patient geometries respectively. The patient geometry can be specified using CTcreate code[14], available with the BEAM/DOSXYZnrc distribution. The multi-leaf collimator (MLC) shapes and motion can be modeled as reported in the literature[8,15-18], with some of these codes being available from the authors. Methods for calibration of a "Monte Carlo linac" and conversion of MC calculated dose (given in Gy/electron) to the absolute dose in Gy have also been described in the literature[19,20]. Coordinate transformations between EGSnrc co-ordinate system and a DICOM-compliant treatment planning system (TPS) co-ordinate system, that are required to re-calculate TPS-produced plans using BEAM/DOSXYZnrc codes, have also been reported by our group recently[21].

An essential component of any automated treatment plan verification process is a capability of extracting all geometry and dosimetric parameters, specific for a treatment plan, from the treatment planning system. Many modern TPS use the DICOM format as a platform for information exchange - importing and exporting treatment plans, CT data and dose distributions. Due to the complicated, recursive structure of DICOM files, extracting the treatment plan parameters such as beam energy, field size, MLC shapes, gantry and couch angles etc. that need to be transferred from the TPS into the MC system is not a trivial task, and often becomes a bottleneck in the development of a MC verification system. Only one example of such software, the DICOM-RT toolbox[22], has been reported so far. The DICOM-RT toolbox is MATLAB based, and therefore comes with extra overhead and potential limitations for the use on different platforms and operating systems.

This paper reviews, in detail, the DICOM tags and structures relevant to MC dose calculations, and reports the development and use of a library (named Vega) of platform-independent, object-oriented C++ codes for dealing with DICOM data. Vega library is a small and succinct piece of software, offering the fundamental functions and classes for reading/modifying/writing DICOM files in a C++ program. This provides the flexibility to extract all MC required data from DICOM files, and write MC produced dose distributions into DICOM files that can then be processed on commercial DICOM-compliant TPS, or in a research treatment planning environment[4].

## 2. Methods and materials

### *2.1. DICOM Data Structure*

DICOM files store information in a recursively defined data structure. At the root of this structure is a sequence of items. Each item has an associated tag (a pair of 4 digit hexadecimal numbers which, for purposes of this paper, will be represented as a pair of numbers enclosed by brackets, delimited by a comma, and with leading zeros), which denotes what kind of information is stored in the item. The information of what each specific tag corresponds to is stored in a dictionary file. Without this information, the DICOM file is just a collection of binary data with no context. Another piece of information associated with an item is a value representation (VR). The VR determines not what is stored in the item, but how it is encoded. The VR can be both explicitly defined in the DICOM file, or be implicit. If the VR is implicit, then information about the VR is obtained by reading the item's tag, and looking up the VR in the data dictionary. Two examples of VR are "IS" (integer string) and "DS" (decimal string). If an item has VR of "IS", then it is a string which represents an integer (e.g. "1729"); if an item has VR of "DS", then it is a string representation of a floating point number (e.g. "-0.618" or "3.00e08"). Ordinary items only contain a specified amount of data, but items with a VR of "SQ" (sequence) are different. These items contain an array of sequences, where a sequence itself contains an array of items. This is where the recursive nature of DICOM files arises. For more information on the DICOM file format, refer to the paper by Riddle and Pickens [23].

## *2.2. DICOM Parameters*

All DICOM files our TPS produced possess an item with tag (0008,0060), called *Modality*, which indicates what kind of information the DICOM file stores. There are four different modalities which contain parameters which then must be extracted in order to run MC calculations. These four modalities are discussed in this section.

### *2.2.1. CT*

DICOM files containing CT information have modality *CT*. The item *Image Position* (0020,0032) contains a string of three decimal values, delimited by backslash characters. The three values specify the x, y, z coordinates (in mm) of the upper left hand corner for this CT slice. *Image Orientation* (0020,0037) contains a string of 6 decimal values, delimited by backslash characters. The first three values specify the x, y, z directional cosines of the first row, and similarly the next three values specify it for the first column. *Rows* (0028,0010) contains a 2 byte unsigned integer that specifies how many rows this CT slice spans, whereas *Columns* (0028,0011) corresponds to the number of columns. *Pixel Spacing* (0028,0030) contains two decimal strings delimited by a backslash. The first value is the distance between the center of each pixel along adjacent rows, and the second along adjacent columns (both in mm). The *Pixel Representation* item (0028,0103). *Pixel Representation* contains a single unsigned 16-bit integer. This is set to 0 if the stored CT values are unsigned integers, or 1 if the stored CT values are signed 2's complement integers. Next, a pair of tags which are important to be aware of are *Rescale Intercept* (0028,1052) and *Rescale Slope* (0028,1053). These each contain a decimal string which defines a linear rescaling of the stored CT data. The stored CT values are contained in the *Pixel Data* item (7FE0,0010), an array of 2 byte integers (*Pixel Representation* determines whether they are signed or unsigned). The CT number associated with each corresponding pixel is then calculated as: (*Rescale Slope*) × (Stored CT value) + (*Rescale Intercept*). The order in which the information is encoded in the *Pixel Data* item is such that the first value corresponds to the upper left pixel (first row, first column). The pixel data is then stored left to right, top to bottom (increasing row, then increasing column).

### *2.2.2. RTSTRUCT*

If the modality of a DICOM file is *RTSTRUCT*, then the file contains structure information. Here we will only discuss DICOM tags and structures relevant to MC dose calculations. When extracting information from this file, the first item that must be identified is *Structure Set ROI Sequence* (3006,0020). This item contains one sequence for each region of interest (ROI). In particular, each sequence contains the two items *ROI Number* (3006,0022) and *ROI Name* (3006,0026). *ROI Number* is an integer string which this current ROI corresponds to, and *ROI Name* is a string which contains the name of the ROI. The next item of interest is *ROI Contour Sequence* (3006,0039), which contains one sequence for each ROI. These sequences each contain a *Referenced ROI Number* (3006,0084) item, which is used to identify which ROI this sequence corresponds to. To determine this, one has to find the sequence of *Structure Set ROI Sequence* which contains the desired ROI name and take the ROI number from this sequence. Then, find the sequence of *ROI Contour Sequence* which contains the same ROI number in the *Referenced ROI Number* item. The contour information is then contained in this sequence under the item *Contour Sequence* (3006,0040). This item contains a listing of sequences, each of which contain data relating to one closed contour, on one plane. Each of these sequences have an item *Number of Contour Points* (3006,0046) containing an integer string denoting the number of points in this contour. The x, y, z coordinates of each of these points are stored in the item *Contour Data* (3006,0050) as x, y, z triplets and written as an array of decimal strings delimited by backslash characters.

*2.2.3. RTPLAN*

DICOM files with modality *RTPLAN* contain all treatment plan parameters, and in particular the parameters required to model radiation beams with BEAMnrc. The first item of interest in the root of the DICOM file is *Fraction Group Sequence* (300A,0070) which contains one sequence. Within this sequence, there are three items which must be processed: *Number of Fractions Planned* (300A,0078), *Number of Beams* (300A,0080), and *Referenced Beam Sequence* (300C,0004). *Number of Fractions Planned* contains an integer string denoting the number of fractions in this plan, and *Number of Beams* is an integer string containing the number of beams in this plan. The item *Referenced Beam Sequence* contains one sequence for each beam in the plan. This sequence contains the items *Beam Meterset* (300A,0086) and *Referenced Beam Number* (300C,0006). *Beam Meterset* contains a decimal string indicating the number of monitor units (MUs) prescribed in the treatment. *Referenced Beam Number* is an integer string which uniquely identifies this beam. The only other item of interest in the DICOM root is *Beam Sequence* (300A,00B0). This item contains a sequence for each beam in the plan. Each sequence has an item *Beam Number* (300A,00C0), which is an integer string that determines which beam the current sequence corresponds to. For plans which contain enhanced dynamic wedges (EDWs), there will also be an item *Wedge Sequence* (300A,00D1) in this sequence. This item contains one or more sequences, the number of which is equal to the integer string *Number of Wedges* (300A,00D0). However, in practice, the number of wedges will usually be one. The sequence(s) of the item *Wedge Sequence* will contain many items, but the only one which is required for the purposes of running MC is *Wedge ID* (300A,00D4). This item contains a string which is a user-supplied identifier for the wedge. For example, one such ID used in our institution is *EDW45OUT*, which corresponds to a 45 degree dynamic wedge where the Y2 jaw is moving.

The last items in the *Beam Sequence* sequence which need attention are *Number of Control Points* (300A,0110) and *Control Point Sequence* (300A,0111). *Control Point Sequence* is an item containing a multitude of sequences, the number of which is stored as an integer string in *Number of Control Points*. Generally speaking, these sequences store information regarding how the geometry of the beam changes over time. This essentially manifests itself as MLC data, and sometimes carriage moves. From here on, each of these sequences will be referred to as control points for simplicity. The control points are indexed from 0 to the total number of control points minus one, with the index of the current control point given by the integer string *Control Point Index* (300A,0112). The first control point is special in that it contains a plethora of non-time dependent information. In particular, the items *Nominal Beam Energy* (300A,0114), *Gantry Angle* (300A,011E), *Beam Limiting Device Angle* (300A,0120), *Patient Support Angle* (300A, 0122), and *Isocenter Position* (300A, 012C) are of interest. *Nominal Beam Energy* is a decimal string storing the energy in MeV that the current beam is configured for; *Gantry Angle* is a decimal string containing the gantry angle of the current beam; *Beam Limiting Device Angle* is a decimal string indicating the angle of the collimator; and *Patient Support Angle* is a decimal string representing the angle of the patient coach. The item *Isocenter Position* consists of three decimal strings delimited by backslash characters. These three values correspond to the x, y, and z position of the isocenter.

The next information that is stored in the first control point is the position of the jaws and the initial position of the MLCs. These are all under the item *Beam Limiting Device Position Sequence* (300A,011A). This item contains multiple sequences, each of which needs to be processed. Each sequence contains the items *RT Beam Limiting Device Type* (300A,00B8) and *Leaf / Jaw Positions* (300A,011C). The string stored in *RT Beam Limiting Device Type* determines what each particular sequence corresponds to. The possible values are *X* or *Y* (for a symmetric jaws), *ASYMX* or *ASYMY* (for asymmetric jaws), or *MLCX* or *MLCY* (for MLCs). The *X* or *Y* character in each of these strings determine in which direction the leaves / jaws are oriented. In the case of jaws, the item *Leaf / Jaw Positions* contains two decimal strings delimited by a backslash character (providing the jaw positions in mm). In the case of MLC data, before processing the data, it is necessary to determine the number of leaves present. Within the current beam sequence, there is an item *Beam Limiting Device Sequence* (300A,00B6) which is similar to *Beam Limiting Device Position Sequence* in that there are multiple sequences contained within, where each one has its role determined by the contained item *RT Beam Limiting Device Type*. In the sequence which corresponds to MLC data, the item *Number of Leaf / Jaw Pairs* (300A,00BC) is of interest. This item contains an integer string which determines the number of leaf pairs that the MLC is comprised of. Once this number is known, the MLC information in *Leaf / Jaw Positions* can be extracted. This item consists of multiple decimal strings delimited by backslash characters. The total number of values present is equal to twice the number of MLC leaf pairs. The values occur in increasing index order, with A leafs first, then B leafs (e.g. 1A, 2A, ... 59A, 60A, 1B, 2B, ... 59B, 60B) and have units of mm. Another item that is present in each control point is *Cumulative Meterset Weight* (300A,0134). This item stores a decimal string that ranges from 0.0 to 1.0. The value for the first control point is always 0.0, and the value for the last control point is 1.0. The value for intermediate control points indicates what fraction of the total prescribed MUs have been delivered so far in the treatment at this point.

The items which need to be parsed from the control points beyond the first are *Control Point Index*, *Beam Limiting Device Position Sequence*, and *Cumulative Meterset Weight*. In the case of a plan with no carriage moves, *Beam Limiting Device Position Sequence* will only contain MLC information for all other control points, but if there is a carriage move present, then jaw information will also be present and the control point which corresponds to the carriage move can be detected by finding the one in which the jaw position changes.

*2.2.4. RTDOSE*

DICOM files containing dose data have modality *RTDOSE*. Such DICOM files have some items in common with CT files: *Image Position* (0020,0032), *Rows* (0028,0010), *Columns* (0028,0011), and *Pixel Spacing* (0028,0030). However, unlike DICOM CT images with one plane per file, DICOM RTDOSE files have all planes stored in a single file. This is manifested in two new items which define the geometry of the dose matrix. *Number of Frames* (0028,0008) is an integer string which defines the number of z-slices in this dose matrix, and *Grid Frame Offset Vector* (3004,000C) is an array of decimal string values delimited by backslash characters, which define the offset of each z-slice from the position defined by *Image Position* (the coordinates of the bottom left pixel on the lowest z-slice). Next, the item *Dose Grid Scaling* (3004,000E) contains a decimal string which is used as a factor to scale the values stored in the *Pixel Data* (7FE0,0010) to Gy. In order to find out which beam number a given radiotherapy (RT) dose file corresponds to, you need to find the item *Referenced Beam Number* (300C,0006) which is deeply nested in the item *Referenced RT Plan Sequence* (300C,0002). *Referenced RT Plan Sequence* contains only one sequence, which contains the item *Referenced Fraction Group Sequence* (300C,0020), which contains only one sequence, which contains the item *Referenced Beam Sequence* (300C,0004), which contains only one sequence, which contains the desired item *Referenced Beam Number*. It is probably the most convoluted way of storing a single integer.

Regardless, armed with knowledge of this number, one can relatively quickly reference the RT plan file to determine what beam this dose file's data corresponds to. Next, before it is possible to interpret the binary data stored in *Pixel Data*, the value stored in *Bits Allocated* (0028,0100) must be read (this value is a 2 byte unsigned integer). If the value is 16, then the dose matrix stored in the item *Pixel Data* is an array of 2 byte unsigned integers, and if the value is 32, then the item is an array of 4 byte unsigned integers. The first value in *Pixel Data* corresponds to the bottom left pixel of the lowest z-slice, and each consecutive value increases left to right, bottom to top, and lowest z-slice to highest z-slice. Finally, to convert the integers stored in *Pixel Data* into units of Gy, you need to read in the decimal string *Dose Grid Scaling* (3004,000E) and multiply each element in the pixel data array by this value.

## 2.3. Vega DICOM Library

In order to extract parameters from DICOM files, software able to deal with DICOM files was developed. The Vega DICOM Library (Vega) is a collection of C++ classes and functions that facilitate the reading, modification, and writing of DICOM format files. The entire DICOM standard encapsulates not just the storing of information, but also communication protocols. Furthermore, the standard places many restrictions on what kind of data is required, optional, or dependent on other data within a DICOM file. For the purposes of Vega, it is not necessary, nor advantageous, to enforce all these specifications. Instead, the Vega library just tries to adhere to the parts of the DICOM standard which specify how data is stored in a DICOM file (sections 5 and 10 of the standard[24]). Also, it is well recognised that many TPSs use internal file formats that although strictly are not required to conform to the DICOM protocol, nevertheless to varying degrees do conform. Therefore, the Vega library was designed to detect when some of the discrepancies encountered during development are present in files it reads in, and will try to correctly read the DICOM file in. For example, some files would not have file preamble and file meta information, whereas others would have the file meta information encoded using the implicit VR little endian transfer syntax instead of explicit VR little endian. If a DICOM file is written using the library, it will aim to ensure that the output file could form part of a DICOM conformant fileset (for instance, it will always write the file meta- information in explicit VR little endian)..Vega also supports all three basic DICOM transfer syntaxes: Implicit VR Little Endian, Explicit VR Little Endian, and Explicit VR Big Endian. It has also been tested on machines with both 32 bit and 64 bit architectures. The library doesn't yet support files containing items or sequences of undefined lengths, because no such files where encountered during development.

Figure 1 shows a diagram of the classes that are contained in the Vega library. CPP_Dictionary is a class which reads in a text file containing all the DICOM tags that Vega can recognize. CPP_Dictionary has an array of objects called CPP_Page, which is a class storing information on each individual tag. Together, these two classes allow the Vega library to understand the kind of information that corresponds to a wide variety of tags. Also, if Vega ever encounters a tag it can't find in its dictionary file, it will output a warning message indicating which unknown tag was encountered. Sometimes, a tag can't be identified because it is a private, developer specific tag, but in many other cases, the tag can be identified through standard DICOM documentation, and simply appended to the dictionary text file, which is provided with the library.

The next important pair of classes are CPP_Sequence and CPP_Item. These are recursively defined classes, which together allow the Vega DICOM library to contain the entire data structure of a DICOM file. At the root of all DICOM files is a sequence of items. This root sequence is an instance of the class CPP_Sequence, and DICOM files are therefore represented in Vega as instances of the CPP_Sequence class. This root CPP_Sequence object then contains an array of CPP_Item objects, corresponding to the items in the root of the DICOM data structure. Most CPP_Item objects then simply contain encoded binary data, which is stored in the Value, which isn't a class, but a union. A C++ union is a data type that, at any given time only contains one object from its list of members. By using a union, it allows Vega to reference the binary data in a variety of contexts. Some items are arrays of characters, others arrays of unsigned 2 byte integers, others arrays of doubles. The context is determined by the item's VR. There are special items, as previously mentioned, that have a VR value of "SQ". In this case, the CPP_Item object doesn't contain a binary data field, but rather an array of CPP_Sequence objects. This is the way in which a DICOM file is represented in the data structure defined by Vega.

In order to extract relevant data from DICOM files, it is first necessary to identify the tags under which the information desired is stored. To do this, a program is provided with the Vega library, called "DICOM_Out", which creates a text-file representation of DICOM tags and data stored within the file. Figure 2 shows an example output of this program, for a chosen RT plan DICOM file. This particular page was chosen as an example because it shows clearly the nested nature of DICOM files. This is the start of a Control Point Sequence, which specifies the positions of a multi-leaf collimator for each step. On each line, there is first an 8 character hexadecimal number corresponding to the tag. Then, if it's an item we then see the two character VR. Also on each line is a string, which is the title the given tag corresponds to. After the title, is the data stored in the item. For instance, item with tag 0x300A011C corresponds to Leaf/Jaw Positions. The data stored in this item is dependent on the value of the item with tag 0x300A00B8, RT Beam Limiting Device Type. If this item is "ASYMX" or "ASYMY", for example, then the data in the Leaf/Jaw Positions item will be the positions of either the X or Y jaws respectively. If it is "MLCX", then the data in the Leaf/Jaw Positions item will be the positions of each leaf in the multi-leaf collimator.

So by using the DICOM_Out program, in correspondence with DICOM documentation, one can determine exactly what is stored in a given DICOM file, and what tags to search for. One then needs only to use this knowledge, and the methods associated with CPP_Sequence and CPP_Item classes to retrieve the data desired. Vega manages to encapsulate all the details of dealing with DICOM files, such that basically all one needs to know are the tags to search for. For example, once one has a handle on the item corresponding to "Control Point Sequence" (tag 0x300A0111), getting the "Beam Limiting Device Position Sequence" (tag 0x300A011A) item for control point index 0 is then a simple call such as:

```
beamLimitingDeviceItem = controlPointItem->seq(0)->item(0x300A011A);
```

Writing to a DICOM file is a considerably more complicated task than reading information from such file. The DICOM standard requires each new DICOM file to have a whole battery of unique IDs (UID) to be generated at the time of creation. Furthermore, if a patient has many DICOM files associated with it (such as plan information, CT information, RT dose distributions), they are all interlocked together through references to their UIDs. As such, creating an entire set of DICOM files is no easy task. Also, since the primary use of writing to DICOM files for our clinic was to store newly calculated Monte Carlo dose distributions in RT Dose DICOM files, it wasn't necessary to create new DICOM files from scratch. Instead, a much easier method is to simply use the original DICOM file exported from the TPS, replace whatever information is required (in our case, dose distributions) while leaving the UIDs untouched, then rewrite this DICOM file. The Vega DICOM library enables the user to write a DICOM file in this way, as simply as they would read in a DICOM file. Simply change all the data using calls to CPP_Item::changeValue(), then call the function CPP_Sequence::DICOM_Write() on the root sequence. Indeed, changing the contents of a DICOM file without changing the UID is a DICOM violation; however, in practice, we only use this function to operate on copies of patient plans that were specifically created for the purpose of Monte Carlo verification. These plan copies are modified slightly to force the TPS to alter their UID while maintaining dosimetrically important parameters of the original plan. This allows the TPS to import DICOM files that contain MC dose distribution, and yet still recognize which CT scans, plan information, and patient they correspond to.

The library is extensively documented using Doxygen, a program for automatic documentation generation from comments stored in the source files. Also included with the library are example codes demonstrating usage of the library, such as the above mentioned DICOM_Out program.

*2.4. Vega Library Validation*

In order to test the library's functionality, RT plans were exported from a commercial TPS[*] as a set of DICOM files and written onto external media. This was performed using the TPS's built-in export utility. These files were processed by several auxiliary programs created using the Vega library, to extract MC required data and prepare the input files required for running BEAMnrc and DOSXYZnrc codes. For the sliding-window IMRT plans, the auxiliary programs extract dMLC files and write them in a standard MLC format; for the conformal plans that use static MLCs the the programs extract the MLC files and re-writes them in the form that models dynamic MLC file. This is for the static MLC files to be compatible with VCU dMLC transport code[17].

Before actually running MC simulations, the treatment plan information extracted from the DICOM files was validated against the values reported by the TPS. Simple values like the jaw positions, the number of MUs per treatment field, gantry angles, collimator angles, couch angles, wedge angles, and isocenter position were verified visually. The MLC sequences extracted were compared against the MLC files which were manually exported through the TPS.

The library has then been integrated into the Vancouver Island Monte Carlo (VIMC) system [25-27], and further verification was done by re-calculating Eclipse treatment plans with VIMC, where all input parameters required for Monte Carlo runs were obtained by Vega library programs. In VIMC, the treatment beams and radiation transport through the linac treatment head and patient are modeled with BEAMnrc and DOSXYZnrc, with the angles of beam incidence extracted from DICOM files and transformed to the DOSXYZnrc coordinate system[21]. IMRT and conformal fields modeled by using VCU dMLC code[17], and enhanced dynamic wedges are modeled as described by Verhaegen and Liu[28]. The MC 3ddose files of each individual field converted then to absolute dose units[20], and placed into DICOM RT dose files. Finally, these dose files, along with the RT plan file, are imported into the TPS for analysis and comparison with the dose distributions calculated by the TPS.

Plans that demonstrate Vega's capability of correctly extracting all parameters and data required for dose verification in a wide range of clinically applicable situations have been selected for these tests. The first plan was designed for seven field head and neck IMRT treatment. This was a "typical" IMRT case, and Vega had to extract all required beam data for each of the fields, including dMLC files. The second plan selected was for nine fields' breast IMRT plan. This plan involved large field size, and in addition to the previous case, it tested Vega's capability to detect carriage moves and correctly extract separate dMLC files for each of these. The last plan, selected to demonstrate Vega's functionality, was prepared for partial breast irradiation. This plan involved static conformal fields with enhanced dynamic wedges (EDWs). It therefore tested Vega's capability of correctly generating dMLC files that model static MLC shapes, rather than dynamic leaf motions as tested in previous plans. This plan also tested Vega's capability of correctly extracting EDW angles and orientations. Correct extraction of variable (non-zero) couch angles was also confirmed in this plan.

## 3. Results

Dose distributions calculated by VIMC system that use Vega library, as well as treatment plan parameters extracted by Vega from selected plans are shown in Figures 3-5, illustrating the functionality of the Vega library.

---

[*] Eclipse$^{TM}$, Varian Medical Systems, Palo Alto, CA.

*Plan 1: Seven field Head & Neck IMRT plan*

Sliding window IMRT plan for head and neck treatment (Figure 3) has been prepared for MC calculation using Vega library. The following parameters of the plan were extracted from treatment plan DICOM files for each beam, and converted to BEAMnrc and DOSXYZnrc input files: beam energy, linac gantry angle, collimator angle, couch angle, beam isocentre position, positions of each jaw, dMLC files, and the number of monitor units.

*Plan 2: Nine field Breast IMRT plan*

The same parameters were extracted from breast IMRT plan (Figure 4) as they were for the head and neck IMRT plan. Additionally, as the fields were large, the Vega library detected MLC carriage moves and extracted MLC files for all sub-fields.

*Plan 3: Accelerated Partial Breast Irradiation plan*

Figure 5 presents the dose distribution for a non-coplanar 3D conformal plan produced for accelerated partial breast irradiation (APBI) treatment. The dose distribution in these planes is conformed to the planning target volume using MLCs, enhanced dynamic wedges (EDW) and a non-coplanar beam arrangement. The Vega library in this case extracted the beam geometry, including gantry, collimator and couch angles (notice non-zero value of the couch angles), jaw positions, and the isocenter position. The library has also identified and extracted the angles and motion direction for EDWs involved in this plan.

Good agreement of the MC dose distributions for these cases, where relatively little tissue inhomogeneity is involved, with the dose distributions produced by commercial TPS confirms correct identification and extraction of the treatment beam parameters.

## 4. Discussion and conclusions

This work describes the Vega library and provides the DICOM information required for extracting treatment plan parameters that are necessary for MC treatment planning calculations.

Results shown in the Figures 3-5 demonstrate utility of the library in MC calculations. DICOM tags for most parameters relevant to MC treatment planning calculations have been identified in this paper and implemented in the Vega library. Exceptions are hard wedges and shielding blocks that are both extremely rarely used in our institution and not implemented in our MC system. The library however does provide the ability to identify DICOM tags and extract the values relevant to these treatment head components. At this time our system only supports photon beams, and extension to support electron beams and applicators is a future project.

Although there are a number of groups that use MC in treatment planning calculations, not much has been published on extracting MC required treatment planning parameters from DICOM files. This is partly because many groups developed Monte Carlo verification systems[2,6,7,29] that are not directly interfaced to a DICOM based TPS and therefore may not require data to be extracted from DICOM files. Spezi et al[22] reported a DICOM-RT toolbox to facilitate MC calculations. The toolbox is Matlab-based and therefore comes with the overhead of using Matlab system. Riddle and Pickens[23] provided succinct description of DICOM structure and tags that are useful for extracting image and acquisition parameters from DICOM files. MC specific parameters are often "hidden" deeply in the DICOM structures as shown in the present paper, and Vega is the first, to our knowledge, compact library written for a freely available C++ compiler that allows easy extracting of MC required parameters from DICOM files. This library has been used extensively at our institution with over a hundred IMRT and conformal, both coplanar and non-coplanar patient plans calculated using its utilities. The library was designed with the aim to be able to handle DICOM data from different planning systems and CT scanners, regardless of the equipment manufacturer. Most extensively the library has been tested in our institution on our TPS with patient images produced by two CT scanners from different manufacturers. A few other institutions in Europe, US and Asia have also obtained a copy of the library, and it is being tested with their applications and equipment. The Vega library can be made available upon request to the authors.

**Acknowledgments**

The authors acknowledge Karl Bush of the University of Victoria, BC, Canada for help in implementation of Vega library to VIMC system. We also acknowledge BCCA Monte Carlo Group for useful suggestions on improving the library utilities.

# Figures

*Figure 1*
Vega DICOM Library class structure

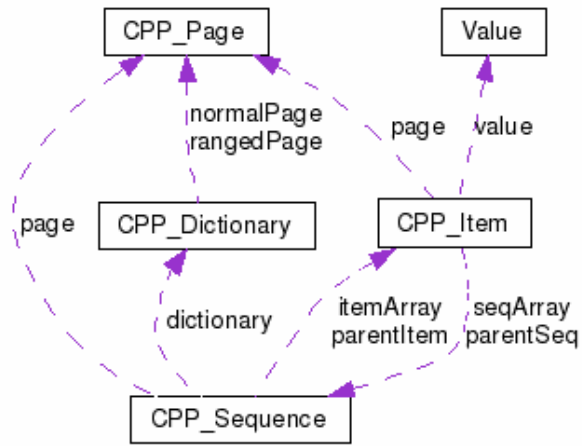

*DICOM library for MC calculations*  13*Figure 2*
DICOM_Out example output

```
0x300A0111 "ControlPointSequence" - Sequence of 20 Items

        0x300A0112 IS "ControlPointIndex": 0
        0x300A0114 DS "NominalBeamEnergy": 6
        0x300A0115 DS "DoseRateSet": 400
        0x300A011A SQ "BeamLimitingDevicePositionSequence" number 1/3

                0x300A011A "BeamLimitingDevicePositionSequence" - Sequence of

                        0x300A00B8 CS "RTBeamLimitingDeviceType": ASYMX
                        0x300A011C DS "LeafJawPositions": -43\48

        0x300A011A SQ "BeamLimitingDevicePositionSequence" number 2/3

                0x300A011A "BeamLimitingDevicePositionSequence" - Sequence of

                        0x300A00B8 CS "RTBeamLimitingDeviceType": ASYMY
                        0x300A011C DS "LeafJawPositions": -110\50

        0x300A011A SQ "BeamLimitingDevicePositionSequence" number 3/3

                0x300A011A "BeamLimitingDevicePositionSequence" - Sequence of

                        0x300A00B8 CS "RTBeamLimitingDeviceType": MLCX
                        0x300A011C DS "LeafJawPositions": -48\-48\-48\-48\-48\-4
8\-48\-48\-48\-48\-48\-48\-48\-48\-48\-48\-48\-48\-48\-48\-48\-48\-48\-
5\-14.9\-48\-48\-48\-48\-48\-48\-48\-48\-48\-48\-48\-48\-48\-48\-48\-48

        0x300A011E DS "GantryAngle": 160
        0x300A011F CS "GantryRotationDirection": NONE
        0x300A0120 DS "BeamLimitingDeviceAngle": 0.0
```

*Figure 3.*

Comparison of seven-field Head & Neck sliding-window IMRT plans calculated using Eclipse TPS (left) and MC (right). All treatment parameters listed in the field details window were extracted from DICOM files and transferred to MC calculations using the Vega library. The library utilities also generated the DICOM dose files that were imported to Eclipse TPS as demonstrated in dose distribution windows.

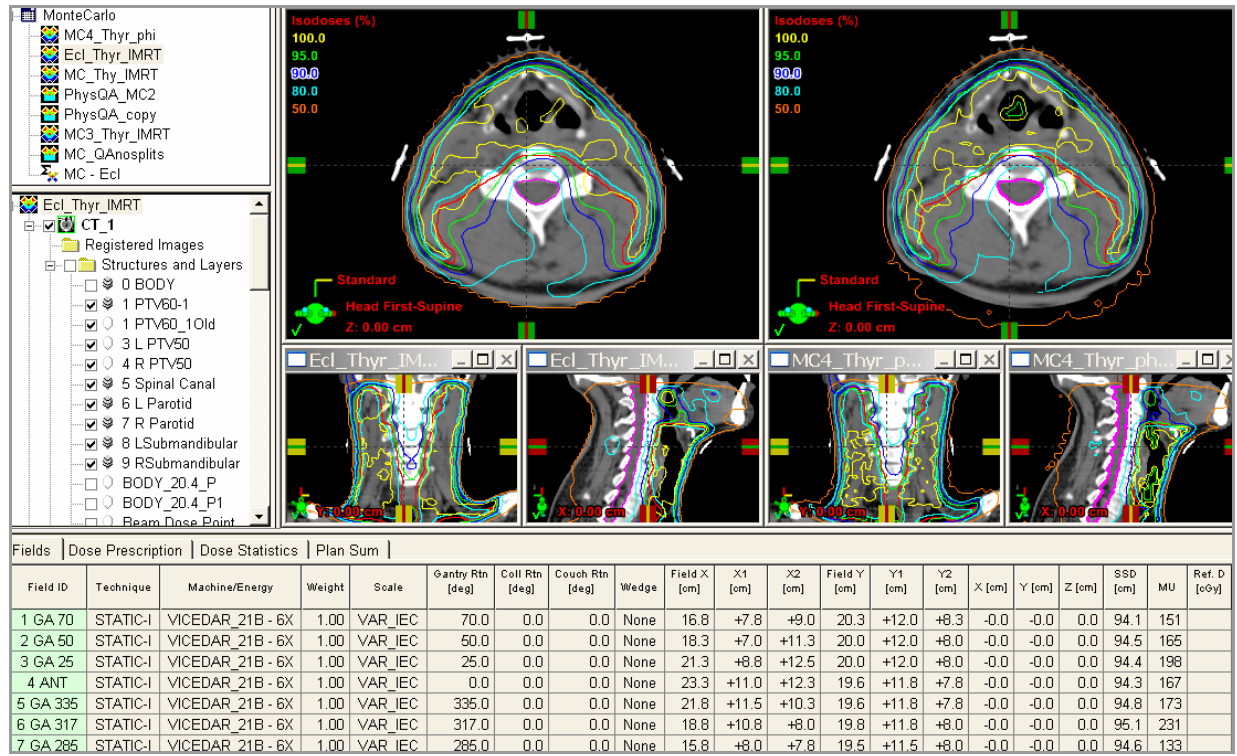

*Figure 4.*

Comparison of nine-field breast sliding-window IMRT plans calculated using Eclipse TPS (left) and MC (right). All treatment parameters listed in the field details window were extracted from DICOM files and transferred to MC calculations using the Vega library. Not seen in the field details are MLC carriage moves that were identified and processed using the library. The library utilities generated the DICOM dose files that were imported to Eclipse TPS as demonstrated in dose distribution windows.

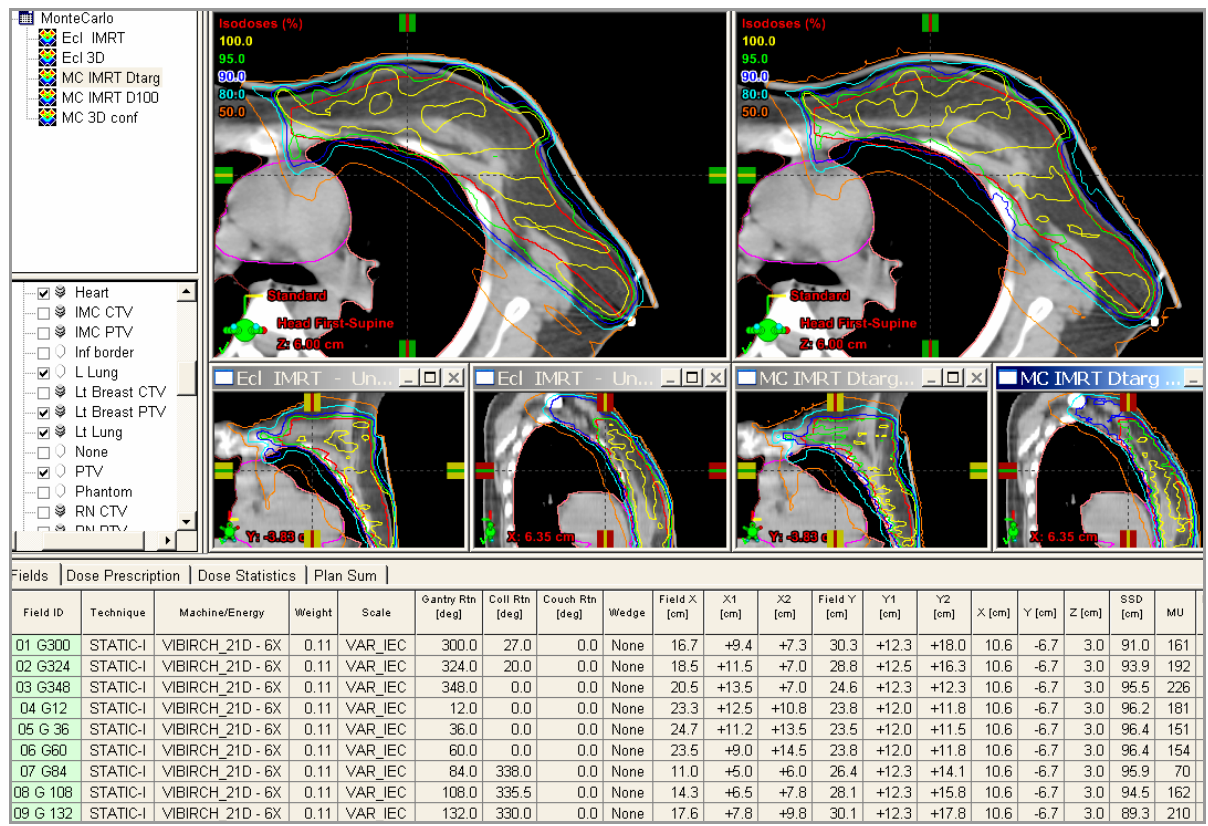

*Figure 5.*

Comparison of four-field non-coplanar conformal APBI plans calculated using Eclipse TPS (left) and MC (right). All treatment parameters listed in the field details window were extracted from DICOM files and transferred to MC calculations using the Vega library. The library utilities also generated the DICOM dose files that were imported to Eclipse TPS as demonstrated in dose distribution windows.

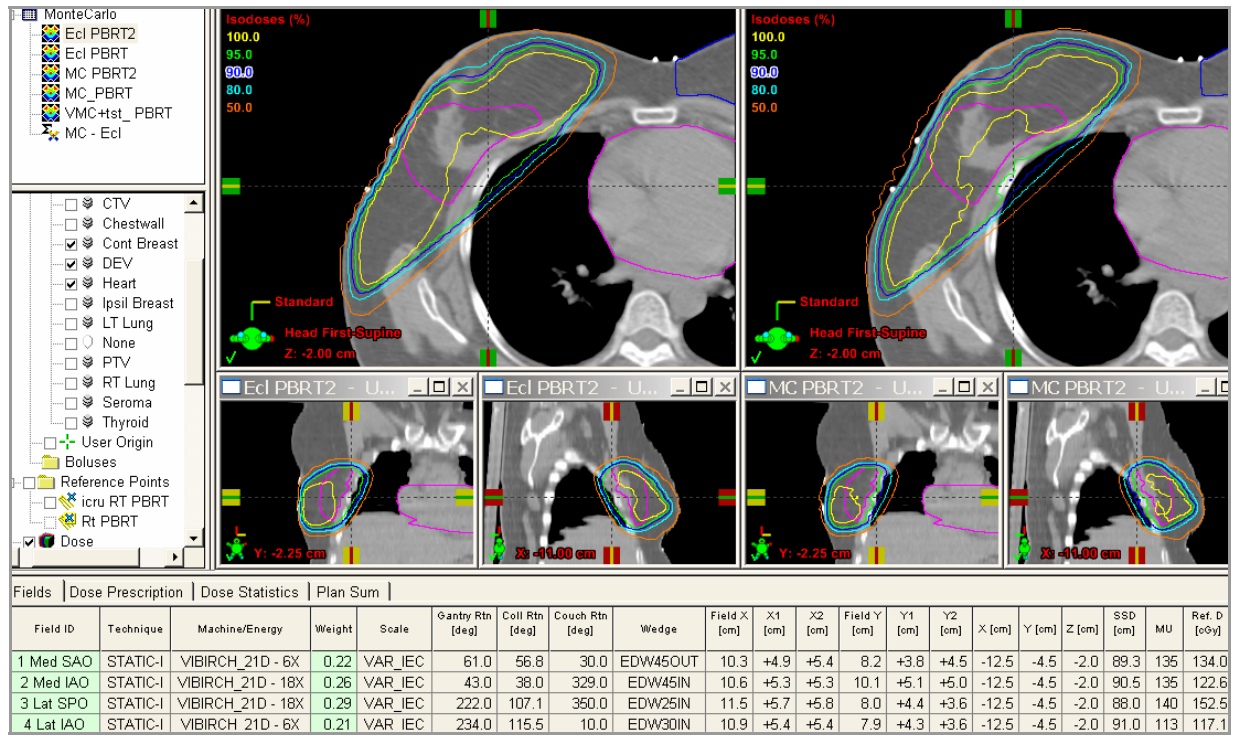